\documentclass{ithaca}

\NUMERO{XII}
\ANNO{2018}

\newcommand{\beq}{\begin{equation}}
\newcommand{\eeq}{\end{equation}}
\newcommand{\bea}{\begin{eqnarray}}
\newcommand{\eea}{\end{eqnarray}}
\newcommand{\lab}{\label}

\def\laq{~\raise 0.4ex\hbox{$<$}\kern 
-0.8em\lower 0.62ex\hbox{$\sim$}~}
\def\gaq{~\raise 0.4ex\hbox{$>$}\kern 
-0.7em\lower 0.62ex\hbox{$\sim$}~}

\def \da {\delta}

\def \b {\beta}
\def \a {\alpha}

\def \Ga {\Gamma}

\def \ep {\epsilon}

\def \r {\rho}
\def \om {\omega}
\def \Om {\Omega}

\def \pa {\partial}

\def \wh {\widehat}

\def \pr {{\prime}}

\def \ra {\rightarrow}

\def\l {\langle}
\def\re {\rangle}

\begin{document}
\title[Onde gravitazionali e fisica moderna]{Le onde gravitazionali nella fisica moderna} 

\dottacitazione{Sembra che Einstein stesso, a un certo punto delle sue ricerche (nel 1936), abbia nutrito qualche dubbio sulla reale esistenza e sulla possibilit\`a di rivelare sperimentalmente queste onde, e che abbia espresso tali dubbi in un articolo che lui stesso sottomise alla famosa rivista {\em Physical Review}. L'articolo, per\`o, fu bocciato dal direttore della rivista, Howard Robinson, e non venne mai pubblicato. In seguito, anche Einstein cambi\`o idea ...}{Aneddoto riportato in un articolo di Faye Flam - Bloomberg, 16 Febbraio 2016}


\author{
\name{Maurizio Gasperini}
\affiliation{Dipartimento di  Fisica, Universit\`a di Bari e INFN, Sezione di Bari, Italy}
}

\ithacamaketitle
\abstract{Una breve introduzione divulgativa alle onde gravitazionali, per rispondere a quelle domande che recentemente un vasto pubblico -- non solo di addetti ai lavori -- si sta ponendo: cosa sono, come si rivelano e, soprattutto, quale impatto ha avuto sulle moderne teorie fisiche la loro recente scoperta? I lettori interessati possono anche trovare alcuni semplici dettagli tecnici e approfondimenti nelle Appendici finali.}


\section{L'antefatto: la fusione di due buchi neri}

Partiamo con l'antefatto: molto tempo fa, in una lontana e sperduta regione del nostro Universo, due corpi celesti molto pesanti e compatti (due ``buchi neri''), si sono venuti a trovare a distanza ravvicinata, si sono attratti gravitazionalmente e si sono fusi insieme. Cosa sono i buchi neri? sono corpi il cui campo di gravit\`a \`e cos\`i forte che nulla, nemmeno la luce, riesce a sfuggirgli (e quindi la loro superficie ci appare completamente nera, opaca). 

Molto tempo dopo, ovvero pi\`u di un miliardo di anni dopo, e precisamente l'11 febbraio 2016, i fisici che lavorano al progetto LIGO annunciano \cite{1} di aver ricevuto e misurato le onde gravitazionali emesse dalla fusione di quei due buchi neri. 

Perch\'e tanto tempo dopo? Perch\'e il processo di fusione e di emissione delle onde si \`e svolto (fortunatamente) a tale distanza dalla Terra che le onde hanno impiegato pi\`u di un miliardo di anni per raggiungerci. Fortunatamente perch\'e, con la distanza, la loro intensit\`a si \`e attenuata: se quell'evento si fosse verificato pi\`u vicino a noi, forse adesso non saremmo qui a parlarne!

\tcbox{Cosa sono le onde gravitazionali?}
{ Sono oscillazioni molto rapide della forza di gravit\`a. Si propagano con la velocit\`a della luce, e trasmettono l'informazione di come il campo gravitazionale di una sorgente cambia nel tempo.}

La rivelazione di tali onde ci porta inevitabilmente ad affrontare una
serie di interessanti domande, che ora enunceremo, e alle quali
cercheremo di dare semplici, ma precise, risposte.

La prima domanda \`e abbastanza scontata.

\section{1. Cosa sono le onde gravitazionali?} 

La risposta \`e riportata nel primo riquadro di questa pagina. 

Tale risposta \`e molto semplice, e ci fa pensare subito alla possibilit\`a di una stretta analogia tra onde gravitazionali (oscillazioni del campo gravitazionale) e onde  elettromagnetiche (oscillazioni del campo elettromagnetico). Tale analogia in effetti esiste, ma \`e meno stringente di quanto potremmo immaginare: ci sono profonde differenze tra i due tipi di onde,  dovute al fatto che l'interazione gravitazionale ammette -- al contrario di quella elettromagnetica --  una importante interpretazione geometrica. 

Nella teoria della relativit\`a generale, infatti, la forza di gravit\`a si pu\`o rappresentare geometricamente come un effetto della curvatura dello spazio (si veda Fig. \ref{fig1}.) Ma se valgono le relazioni 
\begin{center}
\fbox{\footnotesize\texttt{onde $\leftrightarrow$ oscillazioni della forza}}
\end{center}
e inoltre 
\begin{center}
\fbox{\footnotesize\texttt{forza gravitazionale $\leftrightarrow$ curvatura dello spazio}}
\end{center}
ne consegue immediatamente che 
\begin{center}
\fbox{\footnotesize\texttt{onde gravitazionali $\leftrightarrow$ oscillazioni dello spazio}}
\end{center}
Quando il campo di gravit\`a cambia, la forza cambia, e lo spazio si mette a vibrare!

\ColumnFigure{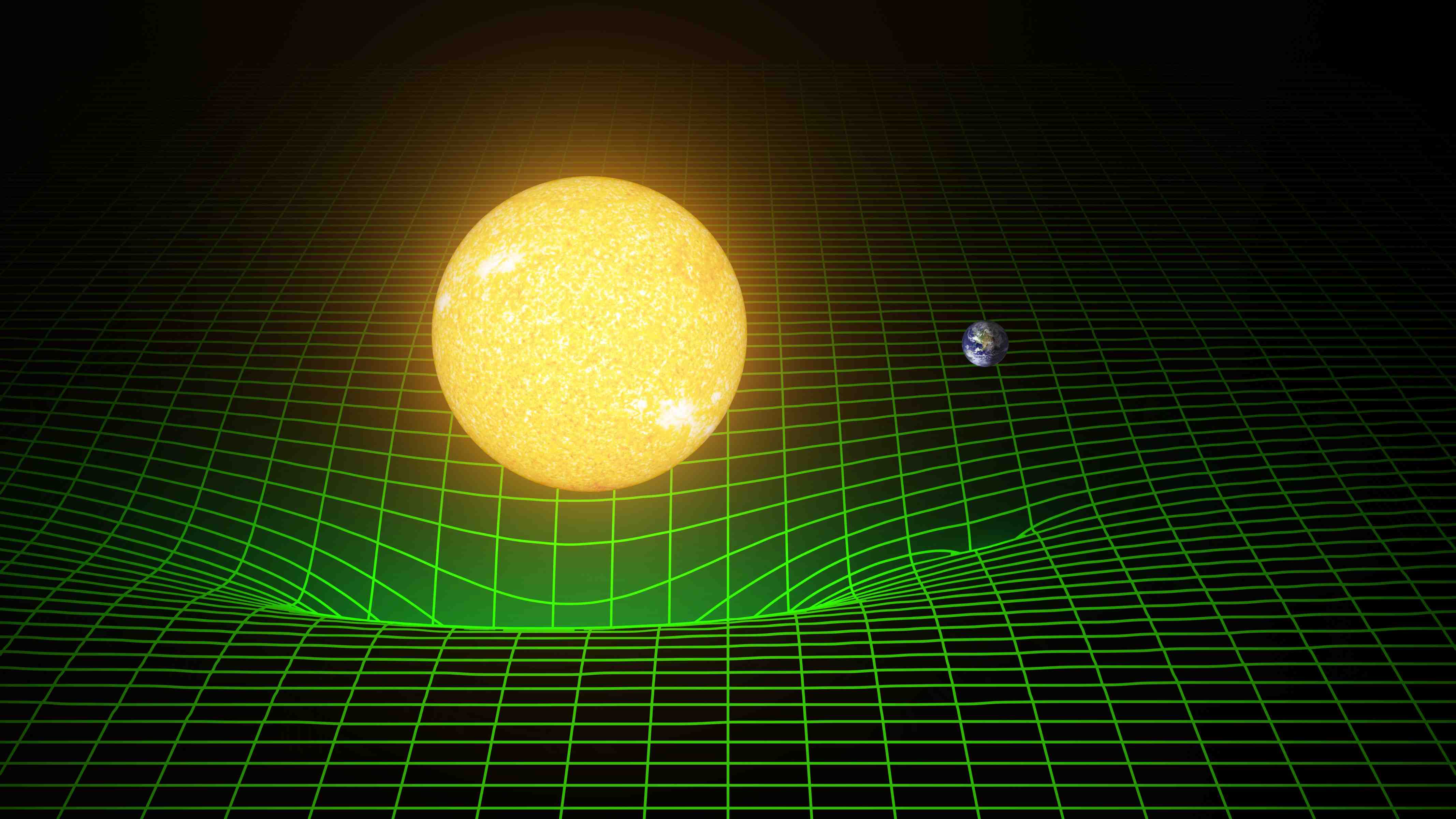}
{\label{fig1}La presenza di massa (pi\`u in generale, di qualunque forma di energia) distorce la geometria euclidea dello spazio vuoto, producendo curvatura in accordo alle equazioni di Einstein.}

Questa interpretazione geometrica delle onde gravitazionali come oscillazioni della geometria spazio-temporale verr\`a brevemente illustrata, con qualche dettaglio tecnico, nelle Appendici 1 e 2. Nella presente Sezione ci limiteremo a riassumere in modo qualitativo alcune conseguenze fisiche di questa interpretazione geometrica, sottolineando le principali differenze tra onde elettromagnetiche (e.m.) e gravitazionali.  

Va detto innanzitutto che, cos\`i come le onde e.m. sono prodotte dal moto accelerato di cariche elettriche, le onde gravitazionali sono prodotte dal moto accelerato delle masse. Per descrivere la radiazione e.m. emessa da un sistema di cariche \`e per\`o sufficiente, in prima approssimazione, considerare l'andamento temporale del momento di dipolo totale del sistema. Nel caso delle onde gravitazionali, invece, ci\`o non \`e possibile. 

Prendiamo ad esempio un sistema isolato composto da un numero finito di cariche elettriche puntiformi, che si muovono con velocit\`a non relativistiche. Il momento di dipolo elettrico del sistema si ottiene moltiplicando ciascuna carica $q_i$ per il corrispondente vettore posizione $\vec{r}_i$, 
e sommando vettorialmente tutti i contributi, 
\[
\vec{d}_\textrm{el} = \sum_i q_i \vec{r}_i
\;\;.
\]
Se la derivata temporale seconda di questo vettore \`e diversa da zero il sistema emette radiazione verso l'esterno, con una potenza (energia per unit\`a di tempo) che \`e data dalla ben nota formula di Larmor. 

Ripetiamo ora la stessa procedura per un sistema isolato di masse puntiformi non relativistiche, e calcoliamo l'analogo vettore di ``dipolo gravitazionale'', sommando vettorialmente i prodotti delle singole masse 
$m_i$ per le corrispondenti posizioni $\vec{r}_i$. 
Il vettore cos\`i ottenuto rappresenta (nell'approssimazione non relativistica) 
il vettore posizione del baricentro $\vec{R}$, 
moltiplicato per la massa totale del sistema $M$: 
\[
\vec{d}_\textrm{gr} = \sum_i m_i \vec{r}_i = M \vec{R}
\;\;.
\]
Ma per un sistema isolato il baricentro si muove di moto rettilineo e uniforme! Quindi la derivata temporale seconda del suo vettore posizione \`e nulla, e quindi non esiste l'analogo  gravitazionale della radiazione di dipolo elettromagnetica (si veda anche l'Appendice 2). 

Come \`e
illustrato nell'Appendice 2, all'emissione di onde gravitazionali contribuiscono, anche in prima approssimazione, non solo le variazioni nel tempo delle velocit\`a ma anche {\em le variazioni nel tempo delle accelerazioni}, che devono essere distribuite in modo sufficientemente ``asimmetrico'' rispetto al centro del sistema. In termini pi\`u tecnici, l'emissione di onde gravitazionali da parte di un sistema fisico isolato \`e controllata, in prima approssimazione, 
dal  momento di quadrupolo $Q_{ij}$ del sistema, che deve essere diverso da zero, $Q_{ij}\not=0$, e che deve variare nel tempo in modo da avere derivata temporale terza diversa da zero, 
\[
\frac{d^3 Q_{ij}} {dt^3} \equiv \stackrel{\dots}{Q}_{ij}\not=0 \;\;.
\]

Ancora: le onde gravitazionali, come  \`e discusso nell'Appendice 1, 
sono caratterizzate da una elicit\`a che \`e 
esattamente il doppio di quella delle onde e.m. . 
L'elicit\`a $\epsilon$ \`e la proiezione, 
lungo la direzione di propagazione $\vec{p} / |\vec{p}|$, 
del momento angolare intrinseco $\vec{s}$ trasportato dall'onda
\[
\epsilon = \frac{\vec{s}\cdot\vec{p}}{|\vec{p}|}
\;\;.
\]
Questa propriet\`a riguarda gli aspetti ``classici'' della radiazione gravitazionale, 
ma ha una importante conseguenza quando tale radiazione viene ``quantizzata'', 
e descritta come un flusso discreto di particelle di massa nulla e di spin (ossia, momento angolare intrinseco) che risulta multiplo intero della costante di Planck 
$\hbar$. L'elicit\`a classica dell'onda gravitazionale implica infatti che i suoi quanti -- i cosiddetti gravitoni -- abbiano spin $s=2$ (in unit\`a di Planck) anzich\`e $s=1$ come i fotoni. Da un punto di vista quantistico la forza gravitazionale si pu\`o quindi pensare come prodotta dallo scambio di particelle di spin 2 e massa zero.

Questo particolare aspetto della forza gravitazionale implica che, per descrivere il campo di gravit\`a a livello classico, sia necessario introdurre un oggetto chiamato ``tensore simmetrico di rango due'', rappresentato da una matrice simmetrica $4 \times 4$ le cui 10 componenti indipendenti descrivono la geometria dello spazio e del tempo. Per descrivere il campo e.m., invece, bastano solo le 4 componenti indipendenti del potenziale vettore e del potenziale scalare (che, combinate insieme, formano il cosiddetto ``quadrivettore potenziale'').

\tcbox{Come possiamo immaginare un'onda gravitazionale?}
{ Come un terremoto! Con la differenza che, per un terremoto, \`e la massa della Terra che vibra; per un'onda gravitazionale, invece, \`e lo spazio stesso che vibra, anche se vuoto.}

\`E importante osservare che queste due differenti descrizioni della gravit\`a e dell'elettromagnetismo non sono arbitrari aspetti ``formali'' delle due teorie, ma sono associate a importanti differenze fisiche tra le due interazioni. Per esempio, al fatto che le forze che agiscono tra masse in moto risultano assai diverse dalle forze tra cariche in moto.  Ma anche nel limite statico, dove le forze sembrano molto simili (variano entrambe con l'inverso del quadrato della distanza), le differenze sono evidenti: la forza dovuta a un campo tensoriale, come quella gravitazionale, produce attrazione tra due masse uguali, mentre la forza dovuta a un campo vettoriale, come quello elettrico, produce repulsione tra due cariche uguali!

Va infine sottolineata, da un punto di vista ``pratico'', l'estrema debolezza delle onde gravitazionali rispetto a quelle elettromagnetiche. Per averne un'idea concreta possiamo calcolare la potenza (ossia l'energia per unit\`a di tempo) della radiazione  emessa da una massa che oscilla in modo armonico con velocit\`a non relativistica. 

In prima approssimazione troviamo che la potenza irraggiata \`e proporzionale alla massa al quadrato, all'ampiezza di oscillazione alla quarta, e alla frequenza alla sesta (si veda l'Appendice 2). La costante di proporzionalit\`a, per\`o, \`e la costante di Newton diviso la quinta potenza della velocit\`a della luce! Un numero troppo piccolo perch\'e le sorgenti di onde gravitazionali costruite in laboratorio possano dar luogo a un segnale rivelabile.

L'esempio fatto suggerisce che per avere un flusso di onde gravitazionali sufficientemente intenso, capace di produrre un segnale nelle antenne che attualmente abbiamo a disposizione, dobbiamo considerare sorgenti di grandi masse e grandi dimensioni come quelle disponibili in un contesto astrofisico -- e, in particolare, come i buchi neri di cui abbiamo parlato all'inizio di questo articolo.

\section{2. Come possiamo immaginarci, e come agisce, un'onda gravitazionale?} 

Per i loro effetti macroscopici possiamo immaginare le onde gravitazionali come le onde d'urto di un terremoto che mettono in vibrazione tutta la materia presente al loro passaggio. Si tratta, per\`o, di un terremoto ``geometrico'', perch\'e \`e lo spazio stesso che vibra, anche se vuoto. Le masse si mettono a oscillare al passaggio di un'onda perch\'e seguono le oscillazioni della geometria nella quale sono immerse.

Questi effetti dell'onda sulle masse (si veda l'Appendice 3 per una discussione pi\`u dettagliata) ci fanno immediatamente capire che i rivelatori di queste onde (le cosiddette ``antenne gravitazionali'') devono funzionare come una specie di sismografi. Ossia, devono mettere in evidenza il movimento delle masse al passaggio dell'onda. Ma se il meccanismo di rivelazione \`e cos\`i semplice, perch\'e \`e risultato cos\`i difficile, fino ad oggi la rivelazione di queste onde?

La risposta \`e fornita dall'intensi\`a di queste onde che, in media \`e estremamente debole (si veda la discussione alla fine dell'Appendice 2). Il pennino di un sismografo geofisico registra spostamenti di qualche millimetro o qualche centimetro. L'antenna gravitazionale, invece, deve registrare spostamenti anche pi\`u piccoli del nucleo di un atomo! Nel caso menzionato all'inizio di questo articolo, ad esempio, il rivelatore LIGO ha misurato spostamenti dell'ordine di $10^{-16}$ cm \cite{1}.

Per misurare spostamenti cos\`i piccoli vengono usati, attualmente, degli interferometri a raggi laser con bracci lunghissimi. Gli interferometri di LIGO, ad esempio, hanno bracci lunghi 4 km. Ma per avere rivelatori ancora pi\`u sensibili si sta progettando di realizzare un interferometro spaziale (LISA) costituito da tre navicelle che, poste in orbita intorno al Sole, darebbero vita a uno strumento 
con bracci effettivi lunghi 5 milioni di km (si veda la Fig.  \ref{fig2}).

Per una presentazione pi\`u dettagliata di queste antenne interferometriche e del loro funzionamento rimandiamo il lettore ad altri articoli contenuti in questo numero di Ithaca dedicato alla gravitazione. 

\ColumnFigure{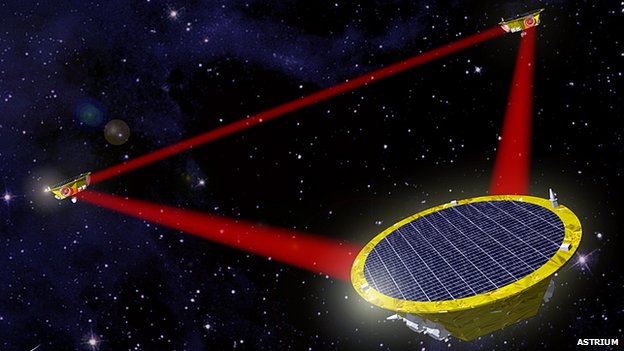}
{\label{fig2}L'interferometro spaziale LISA, progetto della NASA e dell'Agenzia Spaziale Europea (ESA).}

\section{3. Che implicazioni ha per la fisica la rivelazione diretta di queste onde?} 

\tcbox{Una conferma sperimentale \underline{diretta} dell'esistenza dei buchi neri?}
{ I buchi neri sono oggetti talmente compatti da risultare otticamente invisibili perch\`e tutta la loro massa \`e contenuta all'interno di una superficie -- detta ``orizzonte degli eventi'' -- che, perlomeno nel contesto della fisica classica, non lascia trapelare verso l'esterno nessun tipo di segnale o di radiazione (tranne ovviamente la sua attrazione gravitazionale, dovuta alla geometria che si incurva anche fuori dell'orizzonte, si veda la Fig. \ref{fig1}).\\

Quindi, \`e l'effettiva presenza di questo ``schermo'' che distingue un buco nero da un altro oggetto che ha densit\`a simile (o anche maggiore), ma che \`e sufficientemente esteso da evitare la formazione dell'orizzonte. Ricordiamo, a questo proposito, che un corpo di massa $M$ genera un orizzonte solo se il suo raggio risulta inferiore a $2GM/c^2$, dove $G$ \`e la costante di Newton e $c$ la velocit\`a della luce. \\

D'altra parte, quando due oggetti compatti si fondono in un unico oggetto, viene emessa radiazione gravitazionale che si pu\`o scomporre in una serie di modi di frequenza sempre pi\`u elevate. L'ampiezza di tali modi -- soprattutto di quelli emessi nella fase finale del processo, quando il sistema esegue una serie di rapide oscillazioni che smorzano esponenzialmente la sua eventuale anisotropia -- dipende in modo cruciale dalla presenza o meno di un orizzonte degli eventi. \\

Una misura precisa della forma d'onda della radiazione gravitazionale, e delle sue componenti multipolari relative a questa fase finale oscillante, potrebbe confermare (o escludere)  la presenza di un orizzonte degli eventi, e potrebbe quindi confermare (o escludere) in maniera diretta che le onde gravitazionali osservate sono state prodotte da un ``vero e proprio'' buco nero. Le misure attuali sono perfettamente compatibili con questa ipotesi, ma non sono ancora sufficientemente precise da escludere con sicurezza modelli  che prevedono una simile emissione di onde gravitazionali da parte di sorgenti compatte ma senza orizzonte degli eventi.}

Prima di rispondere a questa domanda ricordiamo che si era gi\`a avuta in passato una conferma sperimentale (se pur indiretta) dell'esistenza delle onde gravitazionali,  ottenuta studiando il periodo di rotazione dei sistemi stellari binari \cite{2}. Come verificato dalle accurate osservazioni di Russel Hulse e Joseph Taylor (che per questa scoperta sono stati insigniti del premio Nobel nel 1993), tale periodo decresce nel tempo esattamente come previsto dalla teoria di Einstein se si tiene conto della perdita di energia del sistema dovuta all'emissione di radiazione gravitazionale.

\ColumnFigure{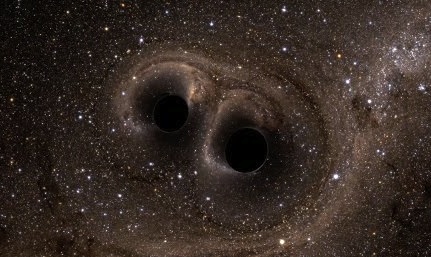}
{\label{fig3}Una simulazione numerica che mostra l'immagine ottica di due buchi neri poco prima della loro fusione.}

I risultati ottenuti dalla collaborazione LIGO-Virgo, a partire dal primo segnale rivelato nel 2015 \cite{1} fino al quarto e pi\`u recente segnale del 2017 \cite{3}, rappresentano invece una vera e propria verifica diretta dell'esistenza delle onde gravitazionali, e forniscono quindi una nuova e importante prova sperimentale della teoria della relativit\`a generale, che si aggiunge ai precedenti e ben noti {\em tests} sperimentali (la deflessione della luce, la precessione del perielio, il ritardo dei segnali radio, etc., si veda ad esempio \cite{4}).

\begin{figure}[t]
\centering
\includegraphics[height=1.3cm]{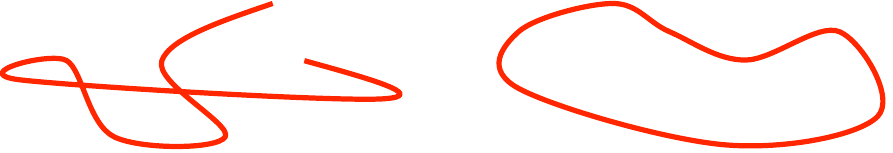}
\caption{Un semplice esempio di stringa aperta e stringa chiusa.}    
\label{fig4} 
\end{figure}

\begin{figure}[t]
\centering
\includegraphics[height=2.6cm]{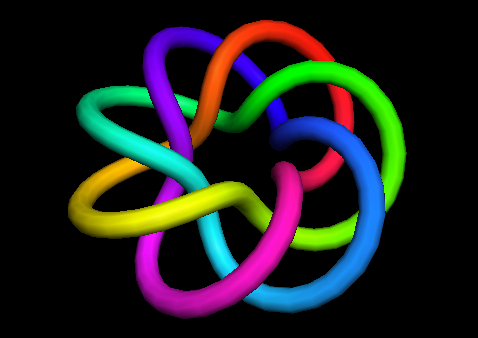}
\caption{Nel contesto della teoria M anche le stringhe acquistano uno spessore, collegato alla presenza della undicesima dimensione.}   
\label{fig5}  
\end{figure}

\tcbox{La teoria delle stringhe}
{
-- \`E un modello teorico secondo il quale i componenti elementari che stanno alla base di tutti i fenomeni fisici osservabili non sono particelle puntiformi ma oggetti estesi unidimensionali (cordicelle, ovvero ``stringhe''), che possono essere aperte o chiuse (si veda la Fig. \ref{fig4}). \\

-- La lunghezza fondamentale di questi oggetti varia a seconda del modello di unificazione considerato ma, in ogni caso,  non pu\`o essere inferiore alla lunghezza di Planck, pari a circa $10^{-33}$ cm.\\

-- Queste microscopiche cordicelle vibrano, e lo spettro di stati quantistici associato a tali vibrazioni riproduce tutte le forme di materia ed energia note (quarks, leptoni, bosoni di gauge), pi\`u una serie infinita di nuove particelle, con  massa e momento angolare intrinseco crescente, ancora da scoprire. \\

-- Per essere consistente con la relativit\`a e la meccanica quantistica il modello deve descrivere una ``superstringa'', ossia un oggetto unidimensionale invariante rispetto allo scambio dei gradi di libert\`a bosonici e fermionici, e deve essere formulato in uno spazio-tempo con 10 dimensioni.\\

-- Esistono solo cinque modelli diversi  di superstringa completi e formalmente consistenti, e tali modelli sono tutti racchiusi in una ``super-teoria'' delle membrane (anche detta ``teoria M''), ambientata in uno spazio-tempo 11-dimensionale (si veda la Fig. \ref{fig5}).}

In aggiunta, la rivelazione di queste onde ci d\`a una prova  (anche se indiretta) dell'esistenza dei buchi neri. Infatti, l'interazione gravitazionale di due buchi neri che porta alla loro fusione, e alla formazione di un buco nero finale (si veda la Fig.  \ref{fig3}), rappresenta il modello teorico pi\`u accreditato (perlomeno al momento) capace di fare da sorgente ai segnali che le antenne gravitazionali hanno captato\footnote{Come discusso nel riquadro di questa pagina, sarebbe bello avere anche osservazioni che forniscono una prova sperimentale {\em diretta} dell'esistenza dei buchi neri. 
Al momento la precisione delle attuali osservazioni non ce lo consente. Speriamo per\`o di ottenere tale conferma, eventualmente, in un futuro non troppo lontano.}.

Ma c'\`e di pi\`u. La rivelazione diretta delle onde gravitazionali, in particolare lo studio delle loro propriet\`a (effettuata mediante l'analisi dei dati forniti dalle antenne), ha importanti implicazioni non solo per la teoria gravitazionale ma anche per una teoria unificata di tutte le interazioni fondamentali: ad esempio, per la teoria delle stringhe (si veda il riquadro presente in questa pagina).  

I risultati sperimentali sulle onde, infatti, ci danno indicazioni positive -- anche se indirette -- sulla possibile esistenza di tale teoria, perch\`e ci confermano che la gravit\`a si comporta come tutte le altre forze riguardo ai processi che regolano l'emissione e la propagazione di radiazione. In particolare, ci confermano che possiamo correttamente associare anche alla radiazione gravitazionale un appropriato tensore energia-impulso, che ne descrive il flusso, la densit\`a d'energia, e che -- soprattutto -- obbedisce alla fondamentale legge di conservazione dell'energia. 

Questi risultati -- che non sono scontati, in principio, data la natura ``geometrica'' dell'energia gravitazionale, e dato il fatto che che la descrizione geometrica varia a seconda dell'osservatore e del sistema di coordinate usato -- ci incoraggiano a pensare che anche la gravit\`a possa essere inclusa in un modello teorico che descrive tutte le forze fondamentali della Natura. 

\WideFigureSideCaption{f6}{\label{fig6}
Un Universo a membrana con 3 dimensioni spaziali e una temporale, immerso in uno spazio esterno che ha 6 dimensioni spaziali ``extra''. Le onde e.m. prodotte da una carica sono confinate sulla membrana. Le onde gravitazionali prodotte da una massa si possono propagare invece sia sulla membrana sia nello spazio esterno multi-dimensionale.}

Esiste un modello di questo tipo? S\`i, il famoso progetto (rimasto incompiuto) di Einstein di costruire una teoria unificata, sembra oggi potersi concretizzare nell'ambito della cosiddetta ``teoria delle stringhe", che rappresenta attualmente l'unico schema capace -- in linea di principio -- di inglobare le interazioni forti, deboli, elettromagnetiche e gravitazionali, e di descriverle in modo consistente sia a livello classico che quantistico, a tutte le scale di distanza e di energia.

Cos'\`e la teoria delle stringhe?  \`E un modello matematico che rappresenta i componenti fondamentali della realt\`a fisica non come oggetti puntiformi (particelle) ma come oggetti unidimensionali estesi (stringhe). Le loro oscillazioni riproducono non solo la gravit\`a ma anche tutte le forze fondamentali della Natura e tutti i componenti elementari della materia. 

Questa teoria rappresenta quindi, a tutti gli effetti, un modello unificato completo, e ha molte conseguenze fisiche rivoluzionarie (si veda \cite{5}), 
quali ad esempio una nuova possibile interpretazione della famosa ``energia oscura'', presente  a livello cosmico
\cite{5a,5b}. Per quel che riguarda la gravit\`a e i suoi aspetti geometrici, in particolare, la teoria delle stringhe prevede che lo spazio abbia pi\`u di 3 dimensioni: devono essercene almeno 9 affinch\`e tale teoria risuti fisicamente (e matematicamente) consistente. Perch\'e allora ci sembra di vivere in uno spazio 3-dimensionale?

Una possibile risposta a questa domanda \`e stata avanzata quasi un secolo fa da Theodor Kaluza e Oskar Klein \cite{6}. Il famodo modello gravitazionale di Kaluza-Klein, che generalizza la teoria di Einstein a uno spazio tempo con 5 (o pi\`u) dimensioni, prevede infatti che le cosiddette dimensioni ``extra'' (ossia, quelle eccedenti le usuali 3 dimensioni accessibili alla nostra quotidiana esperienza) siano `` compattificate'' (ossia, arrotolate strettamente su se stesse), con un raggio di compattificazione cos\`i piccolo da risultare praticamente invisibile a tutti gli attuali esperimenti. 

La teoria delle stringhe suggerisce per\`o anche un'altra possibile soluzione (non necessariamente alternativa alla precedente) al problema dimensionale. La soluzione \`e basata sul cosiddetto modello di Universo ``a membrana'', secondo il quale l'Universo accessibile alla nostra esperienza potrebbe essere solo una ``fetta'' a 3 dimensioni di uno spazio multi-dimensionale. 
In questo modello noi non possiamo percepire le dimensioni in pi\`u, esterne a questa fetta, perch\'e le interazioni fondamentali (che stanno alla base di tutti i nostri sensi e strumenti coi quali esploriamo il mondo circostante) possono propagarsi solo in uno spazio a 3 dimensioni.

C'\`e solo un'eccezione a questa regola, che riguarda da vicino l'argomento di questo articolo: le onde gravitazionali, a differenza degli altri tipi di onde, potrebbero propagarsi anche fuori della membrana (come illustrato in Fig. \ref{fig6}). In questo contesto, per\`o, si producono due importanti conseguenze: $(i)$ l'interazione gravitazionale totale acquista delle componenti massive, a corto raggio d'azione; $(ii)$ la velocit\`a effettiva delle onde pu\`o risultare modificata. 

Il primo effetto potrebbe cambiare il classico andamento della forza gravitazionale di Newton a piccole distanze , introducendo delle correzioni che variano in modo differente dall'inverso del quadrato. Il secondo effetto potrebbe produrre dei ritardi effettivi delle onde gravitazionali rispetto a quelle elettromagnetiche, quando vengono emesse entrambe da una stessa sorgente astrofisica posta a grande distanza. Al momento, per\`o, non abbiamo nessuna evidenza di questi fenomeni n\`e dai {\em tests} sperimentali sulla forza di gravit\`a a piccole distanze, n\`e dal confronto della velocit\`a effettiva delle onde gravitazionali ed elettromagnetiche che \`e stato recentemente effettuato grazie alle osservazioni di LIGO-Virgo \cite{3}, relative a sorgenti che emettono in abbondanza entrambi i tipi di radiazione.

\tcbox{Qual \`e la sorgente pi\`u intensa di onde gravitazionali?}
{
\`E l'Universo stesso! In questo caso, per\`o, l'emissione di onde non \`e direttamente associabile al moto di masse accelerate bens\`i all'accelerazione (o ``inflazione'') della geometria spaziale secondo un meccanismo chiamato ``amplificazione parametrica delle fluttuazioni del vuoto'', studiato per la prima volta da Leonid Grishchuk e Alexei Starobinski negli anni '70. Come funziona questo meccanismo?

 ~~~A livello macroscopico la geometria dello spazio-tempo \`e completamente fissata dalla distribuzione delle masse e delle energie presenti, secondo le equazioni di Einstein. A livello microscopico rimane per\`o una piccolissima indeterminazione della geometria dovuta alla meccanica quantistica, secondo la quale tutti i tipi di campi (e quindi anche quello gravitazionale) possono ``fluttuare'', ossia avere delle piccole oscillazioni locali che li porta a discostarsi per un attimo dal valore assegnato classicamente al campo nel punto dato. 

 ~~~Queste rapidissime variazioni della geometria sono diverse da punto a punto, e possono essere pensate come piccolissime onde gravitazionali virtuali, che non si propagano liberamente ma che sono continuamente emesse e subito localmente riassorbite, e che in media sono nulle. Lo spazio-tempo, secondo la meccanica quantistica, si comporta dunque come un mare che anche quando \`e calmo, e da lontano sembra piatto, visto da vicino presenta invece tante piccolissime ``increspature'' che cambiano continuamente e in maniera casuale. 

 ~~~I quanti delle onde gravitazionali, d'altra parte, sono i gravitoni: questi piccoli disturbi della geometria possono dunque essere pensati come dei gravitoni virtuali, che continuamente si producono, per poi essere immediatamente distrutti. Per non contraddire le leggi di conservazione (ad esempio la conservazione della quantit\`a di moto), questi gravitoni devono per\`o essere prodotti e distrutti in coppia. Se la geometria \`e statica la situazione per le coppie di gravitoni \`e dunque stazionaria: in media, il risultato netto \`e nullo, e il numero medio di gravitoni, nel vuoto, \`e e rimane zero. 

 ~~~Se la geometria varia in maniera sufficientemente veloce (come accade durante il periodo ``inflazionario'')  \`e invece possibile che i gravitoni di una coppia, una volta prodotti, siano trascinati lontano l'uno dall'altro (a causa dell'espansione dell'Universo) in maniera cos\`i rapida da non riuscire pi\`u a ricongiungersi per annichilarsi a vicenda. Rimangono moltissimi gravitoni ``spaiati'', per cos\`i dire, e il risultato netto \`e un abbondante flusso di gravitoni che si distribuisce su tutto lo spazio, riempiendo l'universo in maniera isotropa. 

 ~~~Questo fondo di radiazione gravitazionale, prodotto direttamente dalla geometria in epoche primordiali, pu\`o sopravvivere pressoch\`e indisturbato sino ad oggi grazie alla sua debolissima interazione con il resto delle particelle presenti. Il suo spettro, cio\`e la sua distribuzione in energia, ci trasmette una fotografia fedele della geometria dell'universo all'epoca in cui il fondo \`e stato prodotto. 
}

\section{4. Che implicazioni ci sono per la cosmologia?} 

Le onde recentemente rivelate da LIGO-Virgo \cite{1}, \cite{3} non hanno alcuna implicazione diretta per la cosmologia perch\'e sono state emesse in un'epoca troppo ``recente'' (rispetto alla scala dei tempi cosmici). Infatti, vengono da un'epoca in cui la temperatura media cosmica era solo 0.2 K 
pi\`u alta di adesso! (pi\`u precisamente, era pari a 
2.9  K, mentre il valore attuale \`e 2.7 K). 
La famosa radiazione e.m. ``di fondo''  che si studia in cosmologia \`e stata prodotta invece in epoche molto pi\`u remote, quando l'Universo era circa mille volte pi\`u caldo di adesso\footnote{La radiazione e.m. del fondo cosmico risale all'epoca in cui la materia \`e diventata sufficientemente fredda da smettere di assorbire ed emettere fotoni al ritmo giusto per restare in equilibrio termico, cosicch\`e tale radiazione ha potuto propagarsi liberamente fino ai giorni nostri. A quell'epoca l'universo aveva una temperatura media di circa 2973 K.}  (si veda ad esempio \cite{7}, Cap. 7.3). 

Se andiamo ancor pi\`u indietro nel tempo, per\`o, troviamo che l'Universo diventa cos\`i denso  da non essere pi\`u trasparente alla radiazione e.m., mentre possiamo pensare che resti trasparente alla radiazione gravitazionale fino alla cosiddetta epoca di Planck, ossia alla fase cosmologica in cui l'Universo \`e caratterizzato da un raggio di curvatura dell'ordine della lunghezza di Planck, definita da $\sqrt{\hbar G/c^3}$, e in cui la temperatura cosmica \`e data dalla temperatura di Planck, definita da $\sqrt{\hbar c^5/k^2 G}$ e pari a circa $10^{32}$ K
 (ricordiamo che $\hbar$ \`e la costante di Planck, $c$ \`e la velocit\`a della luce, $G$ \`e la costante di Newton e $k$ \`e la costante di Boltzmann).  Per raggi pi\`u piccoli e temperature pi\`u elevate si entra necessariamente nell'ambito della 
 gravit\`a quantistica, e non \`e pi\`u possibile applicare un modello cosmologico classico (o semiclassico) come quello che stiamo considerando. 

In ogni caso, sono le onde gravitazionali -- e non quelle e.m. -- che possono darci informazioni {\em dirette} sulla storia pi\`u remota del nostra Universo, e in particolare sulle epoche immediatamente successive (o precedenti) il Big Bang. La domanda ovvia, a questo punto, \`e allora la seguente: esistono sorgenti gravitazionali cos\`i potenti da aver prodotto segnali che, pur essendo cos\`i remoti nel tempo, sono ancora abbastanza intensi da essere accessibili alle nostre osservazioni?

La risposta \`e positiva: una potentissima sorgente di onde gravitazionali primordiali esiste ed \`e rappresentata dall'Universo stesso, il quale, durante la sua evoluzione iniziale, pu\`o produrre un fondo cosmico di radiazione grazie al meccanismo illustrato nel riquadro della pagina successiva. I resti fossili di questo fondo sono tutt'ora presenti e ci trasmettono una foto -- presa ``dal vivo'' -- delle era cosmiche ``preistoriche''. 

Diventa dunque molto importante studiare le propriet\`a di questo fondo e chiedersi, in particolare, se la sua intensit\`a pu\`o produrre segnali rivelabili nella banda di frequenza alla quale sono sensibili le attuali antenne.

La risposta a questa domanda dipende fortemente dalle propriet\`a spettrali
del fondo (ovvero, dalla distribuzione della sua energia nelle varie bande di frequenza); tali propriet\`a, a loro volta, dipendono dalla dinamica della fase cosmologica responsabile della sua produzione. \`E allora conveniente, in questo contesto, caratterizzare il fondo di onde gravitazionali mediante la cosiddetta ``densit\`a spettrale di energia'', $\Om_g$, definita come segue. 

Si scompongono (mediante analisi di Fourier) le onde gravitazionali che contribuiscono al fondo cosmico in tutte le loro componenti di frequenza $\om$; per ogni valore di $\om$ si calcola la densit\`a d'energia $\r(\om)$ corrispondente; si esprime infine tale densit\`a in unit\`a di densit\`a cosmologica critica, $\r_c(t)$ (ossia  quella densit\`a di energia che, a ogni  tempo $t$ dato, rappresenta la sorgente nelle equazioni di Einstein che descrivono un modello d'Universo omogeneo e isotropo con curvatura spaziale nulla). In questo modo si arriva alla seguente definizione di densit\`a spettrale: 
\beq
\Om_g(\om, t)= {\om\over \r_c(t)} {d \r(\om) \over d\om}.
\nonumber
\eeq
Possiamo ora distinguere tre principali situazioni. 

Consideriamo innanzitutto il fondo che corrisponde ai resti fossili del Big Bang termico, ossia dell'esplosione di radiazione che pone fine all'epoca di espansione accelerata (inflazione), e segna l'inizio della fase di evoluzione standard (si veda il riquadro sul Big Bang della pagina successiva). Per valori della temperatura cosmica inferiori alla temperatura di Planck (citata in precedenza) la radiazione gravitazionale prodotta da questa esplosione primordiale non viene assorbita dalla altre forme di materia presenti a livello cosmico, ma evolve in modo indipendente, e la sua densit\`a di energia rimane distribuita in frequenza con uno spettro termico (di corpo nero), di tipo Planckiano (del tutto simile allo spettro della attuale radiazione cosmica e.m.).

\tcbox{Cosa si intende esattamente con ``Big Bang''?}
{-- Nel contesto del modello cosmologico standard Big Bang \`e sinonimo della singolarit\`a iniziale che \`e presente nelle equazioni classiche della relativit\`a generale, che descrive uno stato cosmico di curvatura infinita, densit\`a infinita, temperatura infinita, e che segna l'origine (invalicabile) dello spazio e del tempo.\\

-- Come in tutti i modelli fisici in cui le singolarit\`a classiche vengono regolarizzate dalle correzioni quantistiche, anche in cosmologia \`e possibile formulare scenari in cui, andando indietro nel tempo, la curvatura e la densit\`a non diventano infinite ma, raggiunto un valore massimo (comunque molto elevato), hanno una specie di ``rimbalzo'' (detto {\em bounce}) e ritornano a decrescere, pi\`u o meno lentamente. In questo contesto il Big Bang indica una fase di transizione (che pu\`o essere comunque di tipo violento ed esplosivo), che segna il passaggio da un regime cosmico iniziale (a curvatura crescente) al regime finale (curvatura decrescente) descritto dal modello cosmologico standard.\\

-- Infine, c'\`e un terzo possibile significato della parola Big Bang che recentemente sta diventando sempre pi\`u consueto. In tutti i modelli cosmologici realistici (di tipo standard o non standard), la fase di espansione iniziale accelerata (detta ``inflazione'', e necessaria per produrre le disomogeneit\`a che danno luogo alle strutture cosmiche e alle anisotropie osservate) produce un effettivo raffreddamento globale dell'Universo, e va quindi necessariamente seguita da una successiva epoca di riscaldamento, in cui si produce la radiazione termica che oggi osserviamo. Questa calda esplosione di radiazione finale viene chiamata ``{\em reheating}'' (riscaldamento), o anche ``Big Bang termico''. Pu\`o coincidere (ma non necessariamente) col Big Bang che descrive la transizione (o rimbalzo) della curvatura. \\

-- Nel testo di questo articolo facciamo
riferimento alle onde gravitazionali associate a questi diversi tipi di Big Bang.
}

Dal confronto col fondo di radiazione e.m. che tutt'oggi osserviamo possiamo allora dedurre, per queste onde gravitazionali, un picco dello spettro collocato attualmente attorno a una frequenza $\om_\textrm{Max}$ di qualche centinaio di GHz, e una attuale intensit\`a  di picco, $\Om_g(\om_\textrm{Max}, t_0)$, che pu\`o arrivare al massimo a circa un decimo di quella elettromagnetica (che \`e data da $\Om_{e.m.}(t_0) \sim 10^{-4}$): quindi $\Om_g(\om_\textrm{Max}, t_0) \laq 10^{-5}$. Si noti che valori pi\`u elevati di $\Om_g$ rispetto a $\om_{e.m.}$ sono proibiti, perch\`e sarebbero in contrasto coi dati che abbiamo sul processo di formazione primordiale degli elementi leggeri (nucleosintesi), che ha avuto luogo in passato quando l'Universo era sufficientemente caldo; ciononostante, il valore di picco permesso per questo tipo di fondo gravitazionale rimane alto. 

A causa del suo andamento termico, per\`o, lo spettro decresce molto rapidamente 
al diminuire della frequenza (infatti $\Om_g(\om)$ varia con la frequenza in modo proporzionale a $\om^3$ per $\om < \om_\textrm{ Max}$), 
e quindi la sua intensit\`a diventa troppo debole per essere rivelabile 
nella banda di sensibilit\`a degli attuali interferometri (che \`e centrata intorno ai 10-100 Hz, sia per LIGO che per Virgo).  

\WideFigureSideCaption{f7}{\label{fig7}
Esempi di spettro (in scala logaritmica) per un fondo cosmico di onde gravitazionali prodotto in fasi primordiali precedenti (spettro crescente) o successive (spettro decrescente) al Big Bang (dove con Big Bang si intende l'epoca in cui l'Universo raggiunge lo stato di massima curvatura). La figura mostra anche i limiti sperimentali imposti dall'anisotropia cosmica, dalle {\em pulsars}, dalla nucleosintesi,  e dai recenti dati forniti dalle antenne LIGO-Virgo.}

Diversa \`e la situazione per un fondo di onde gravitazionali che viene prodotto prima del Big Bang termico, durante quella fase denominata ``inflazione'' e caratterizzata da una veloce espansione accelerata della geometria cosmica. In questo caso possiamo distinguere due situazioni complementari, a seconda che la curvatura e la densit\`a media dell'Universo, durante l'inflazione, tendano a crescere o a decrescere nel tempo. Ovvero -- facendo ancora riferimento al riquadro sui vari significati del termine Big Bang -- a seconda che la fase di inflazione si collochi, rispettivamente, {\em prima} o {\em dopo} l'epoca del Big Bang di transizione (o di rimbalzo).

Cominciamo col considerare lo scenario pi\`u convenzionale in cui non c'\`e nessuna transizione o rimbalzo, oppure, anche se c'\`e, la fase inflazionaria si svolge comunque in epoche successive, quando l'Universo aveva una curvatura media costante o leggermente decrescente nel tempo. In questo caso il fondo di onde gravitazionali prodotto \`e caratterizzato da uno spettro, detto 
di Harrison-Zeldovich, che risulta pressoch\`e costante (o leggermente decrescente) con la frequenza. I segnali di questo fondo sono dunque accessibili all'osservazione, in principio, anche nel regime di basse frequenze, dove sono sensibili le attuali antenne gravitazionali. 

L'intensit\`a di questo spettro, in principio arbitraria, risulta per\`o fortemente limitata dalle recenti misure riguardanti l'anisotropia del fondo e.m. di micro-onde (si veda la Fig. \ref{fig7}). I vincoli imposti da tali misure -- effettuate nel corso degli anni dai satelliti COBE, WMAP, PLANCK con precisione via via crescente -- forniscono un limite superiore molto stringente per l'intensit\`a di uno spettro gravitazionale di tipo Harrison-Zeldovich: 
il suo valore massimo, nella banda di sensibilit\`a delle attuali antenne interferometriche (ossia, per frequenze dell'ordine di 10-100 Hz) deve soddisfare la condizione $\Om_g \laq 10^{-14}$. Questa intensit\`a \`e troppo bassa per prevedere una rivelazione diretta da parte degli attuali rivelatori (ma ci sono progetti che, in un futuro distante qualche decennio, potrebbero raggiungere la sensibilit\`a necessaria). 

Rimane infine da considerare lo scenario in cui l'epoca inflazionaria precede nel tempo il momento del ``rimbalzo'' e del passaggio alla fase di curvatura decrescente. In questo caso il Big Bang di transizione pu\`o anche coincidere col Big Bang termico (si veda il riquadro sul Big Bang della pagina precedente). In ogni caso, l'inflazione ha luogo durante un regime cosmico molto diverso da quello standard, che ha la funzione di ``preparare'' lo stato attuale dell'Universo portando la densit\`a e la curvatura ai valori richiesti, instaurando la giusta gerarchia tra le costanti di accoppiamento delle varie interazioni, rompendo alcune simmetrie e restaurandone altre, etc. 

Ma quello quello che \`e importante -- per l'oggetto di questo articolo -- \`e che l'inflazione ha luogo durante una fase in cui la curvatura media dell'Universo cresce nel tempo: di conseguenza, produce un fondo cosmico di onde gravitazionali il cui spettro tende a crescere con la frequenza (ma non cos\`i velocemente, in generale, come lo spettro termico). 

Con uno spettro di questo tipo l'intensit\`a del fondo pu\`o risultare sufficientemente debole a basse frequenze -- cos\`i da rispettare i limiti osservativi imposti dall'anisotropia cosmica (ed anche i limiti superiori imposti dalle {\em pulsars}, si veda ad esempio \cite{7}, Cap. 8.3.1) -- e ciononostante avere un'intensit\`a sufficientemente alta nella banda di sensibilit\`a delle antenne  (si veda la Fig. \ref{fig7}) -- cos\`i da risultare rivelabile in un futuro non troppo lontano. Il vincolo pi\`u stringente, per questo tipo di spettro, \`e fornito dai processi di formazione primordiale degli elementi (nucleosintesi), che sarebbero disturbati dalla presenza di onde gravitazionali troppo intense, e che impongono la condizione $\Om_g \laq 10^{-5}$ sul valore di picco dello spettro (come abbiamo gi\`a sottolineato in precedenza).

I risultati di questa discussione sono illustrati graficamente dalla Fig. \ref{fig7}, dove si presentano alcuni espliciti esempi di distribuzioni spettrali di tipo crescente e decrescente prodotti, rispettivamente, da modelli di evoluzione inflazionaria con curvatura cosmica crescente o decrescente. Il picco degli spettri crescenti \`e stato fissato a $\Om_g=10^{-6}$ e $\om_\textrm{ Max}=10^{11}$ Hz (che corrispondono ai valori massimi suggeriti dalla teoria delle stringhe). Tutti gli spettri presenti in figura risultano compatibili sia coi vincoli sperimentali indiretti (imposti da anisotropia cosmica, {\em pulsars}, nucleosintesi) sia coi vincoli {\em diretti} imposti dalle pi\`u recenti osservazioni di LIGO-Virgo \cite{8}, che implicano $\Om_g \laq 5 \times 10^{-8}$ alla frequenza $\nu= \om/2\pi= 25$ Hz. 

\`E interessante osservare, in questo contesto, che la sensibilit\`a sperimentale delle antenne LIGO-Virgo \`e ancora in fase di miglioramento, e che nel 2020 arriver\`a al livello di sensibilit\`a corrispondente a  $\Om_g=10^{-9}$. Ancora pi\`u elevata \`e la sensibilit\`a prevista dell'interferometro spaziale LISA (Fig. \ref{fig2}), progettato per raggiungere il livello 
 $\Om_g=10^{-13}$ (per\`o a frequenze pi\`u basse, $\om \sim 10^{-2}$ Hz). 
 
 Possiamo quindi concludere che, anche rispettando tutti i vincoli noti, non \`e escluso che un fondo cosmico di onde gravitazionali primordiali possa essere direttamente osservato in un prossimo futuro (portandoci i segnali di un Universo veramente ``preistorico''), a patto che il suo spettro cresca opportunamente con la frequenza, come illustrato in Fig. \ref{fig7} (si veda \cite{9} per una discussione dettagliata di questa possibilit\`a).

\section{Appendice 1. Le oscillazioni della geometria} 

In questa Appendice mostreremo che le onde gravitazionali si propagano nel vuoto alla velocit\`a della luce, sono caratterizzate da due stati di polarizzazione linearmente indipendenti, e hanno elicit\`a pari a 2.

Consideriamo per semplicit\`a il caso in cui i campi gravitazionali in gioco 
siano sufficientemente deboli, e possano essere descritti dalle equazioni di Einstein linearizzate. Supponiamo che la metrica dello spazio-tempo, $g_{\mu\nu}$, si discosti poco da quella piatta di Minkowski, $\eta_{\mu\nu}$, e poniamo 
\beq
 g_{\mu\nu} \simeq \eta_{\mu\nu}+h_{\mu\nu} ,
 \lab{a1}
\eeq
dove il tensore simmetrico $h_{\mu\nu}$ soddisfa la condizione $|h_{\mu\nu}| \ll 1$, e descrive delle piccole distorsioni locali della geometria spazio-temporale prodotte dalla presenza del campo gravitazionale. 

\tcbox{Sulla velocit\`a delle onde gravitazionali}
{Nel vuoto (e in una geometria piatta) \`e pari alla velocit\`a delle onde elettromagnetiche. Ma pu\`o essere diversa dalla velocit\`a della luce in presenza di un mezzo, o in presenza di un campo gravitazionale sufficientemente intenso da rendere necessarie alcune correzioni alla teoria gravitazionale di Einstein (che rimane valida in ambito classico e macroscopico). Ad esempio, correzioni in serie di potenze della curvatura e della costante di accoppiamento (come previsto dalla teoria delle stringhe).}

In questa approssimazione trascuriamo i termini di ordine quadratico (o superiore) in $h$, per cui le componenti covarianti e controvarianti di $h$ sono collegate tra loro dalla metrica di Minkowski, 
\beq
h_\mu\,^\nu= g^{\nu\a}h_{\mu\a} =\eta^{\nu\a}h_{\mu\a}+ {\cal O}(h^2),
\lab{a2}
\eeq
e la metrica inversa (tale che $g^{\mu\a} g_{\nu\a} = \da^\mu_\nu$)  \`e data da
\beq
g^{\mu\nu}\simeq \eta^{\mu\nu}-h^{\mu\nu}. 
\label{a3}
\eeq
Le componenti della connessione di Christoffel, in questo contesto, assumono la forma approssimata
\beq
\Ga_{\nu\a}\\^\b\simeq{1\over 2}\left( \pa_\nu h_{\a}\,^\b
+\pa_\a h_{\nu}\,^\b-\pa^\b h_{\nu\a}\right), 
\label{a4}
\eeq
e il tensore di Ricci si riduce a 
\bea
&& 
R_{\nu\a}= R_{\mu\nu\a}\,^\mu \simeq
\pa_\mu \Ga_{\nu\a}\,^\mu- \pa_\nu \Ga_{\mu\a}\,^\mu =
\nonumber \\ &&
={1\over 2} \big( \pa_\mu\pa_\a h_\nu\,^\mu - \Box h_{\nu\a} -  \pa_\nu\pa_\a h+ 
\pa_\nu\pa_\mu h^\mu\,_\a\big), 
\nonumber \\ &&
\label{a5}
\eea
dove $h=h_\mu\,^\mu$  \`e la traccia del tensore $h_{\mu\nu}$, e dove  $\Box$  \`e l'usuale operatore D'Alembertiano dello spazio piatto,
\beq
\Box= \eta^{\mu\nu} \pa_\mu\pa_\nu= \nabla^2-{1\over c^2} {\pa^2\over \pa t^2}.
\label{a6}
\eeq

Sfruttiamo ora le simmetrie della relativit\`a generale, e in particolare l'invarianza dell'azione per trasformazioni locali e infinitesime di coordinate, che ci permette sempre di imporre  sulla metrica 4 arbitrarie ``condizioni di {\em gauge}''. Nel  nostro caso \`e conveniente scegliere il cosiddetto {\em gauge} armonico, utilizzando un sistema di coordinate dove la metrica (\ref{a1}) soddisfa la condizione 
\beq
\pa_\mu h_\nu\,^\mu={1\over 2} \pa_\nu h.
\label{a7}
\eeq
Usando questa relazione nell'Eq. (\ref{a5}) troviamo infatti che le componenti del tensore di Ricci si semplificano notevolmente, e le equazioni di Einstein nel vuoto ($R_{\nu\a}=0$) si riducono alla classica equazione d'onda
\beq
\Box h_{\nu\a}=0,
 \label{a8}
\eeq
la cui soluzione descrive segnali che si propagano nel vuoto con la velocit\`a della luce. I segnali, in questo caso, sono le fluttuazioni della geometria descritte dalle componenti del tensore $h_{\mu\nu}$.

Le componenti indipendenti del tensore simmetrico $h_{\mu\nu}$ sono in generale 10, ma si riducono a 6 dopo aver imposto le 4 condizioni del  {\em gauge} armonico (\ref{a7}). Vedremo ora che -- con un'opportuna scelta del sistema di coordinate che non ci fa uscire dal  {\em gauge} armonico -- possiamo sempre imporre 4 ulteriori condizioni su $h_{\mu\nu}$, che riducono a 2 il numero totale delle sue componenti indipendenti.

Questo significa che il sistema di equazioni (\ref{a7}), (\ref{a8}), che descrive la propagazione di onde gravitazionali nel vuoto, \`e caratterizzato da due soli gradi di libert\`a dinamici (che corrispondono fisicamente ai due possibili stati di polarizzazione dell'onda). Facciamo anche vedere che \`e sempre possibile scegliere il sistema di coordinate in modo tale che $h_{\mu\nu}\not=0$ solo per le componenti spaziali ortogonali alla direzione di propagazione dell'onda. 

Supponiamo ad esempio che l'onda si propaghi lungo la direzione dell'asse $x$, e sia quindi descritta da una generica soluzione ritardata dell'equazione di D'Alembert, del tipo $h_{\mu\nu}(x,t)= h_{\mu\nu}(x-ct)$. In questo caso la condizione di  {\em gauge} (\ref{a7}) si riduce a
\beq
\pa^0 h_{\mu 0}+ \pa^1 h_{\mu1}= {1\over 2} \pa_\mu h,
\label{a9}
\eeq
dove $\pa_0$ indica la derivata rispetto al tempo e $\pa_1$ quella rispetto a $x$.  Introduciamo un nuovo sistema di coordinate, definito dalla trasformazione infinitesima $x^\mu \ra x^{\mu}+ \xi^\mu(x) + \cdots$, e determiniamo $\xi^\mu$ (il generatore della trasformazione) imponendo nel nel nuovo sistema valga ancora la condizione di {\em gauge} (\ref{a9}) e, in pi\`u, sia soddisfatta la nuova condizione
\beq
h_{\mu0}=0.
\label{a10}
\eeq
La trasformazione cercata si ottiene facilmente calcolando la variazione locale infinitesima del tensore metrico, e imponendo le condizioni richieste (si veda ad esempio \cite{4}, Cap. 9.1.1).

\tcbox{I gravitoni}
{Le onde gravitazionali si propagano lungo il cono luce dello spazio-tempo di Minkowski, rappresentano una sovrapposizione di due stati di polarizzazione linearmente indipendenti, e le onde polarizzate circolarmente hanno elicit\`a $\pm 2$. Ne consegue che i quanti dell'onda gravitazionale -- i gravitoni -- devono avere massa zero e spin $s=2$,  orientato in modo parallelo o antiparallelo alla direzione di moto.}

In queste nuove coordinate la condizione di {\em gauge} (\ref{a9}) si riduce a 
\beq
\pa^1  h_{\mu1}= {1\over 2} \pa_\mu h,
\label{a11}
\eeq
dove tutte le componenti di $h_{\mu\nu}$ con indici uguali a  zero sono nulle. Ponendo $\mu=0$ abbiamo allora $\pa_0h=0$, ossia $h=$ costante, il che significa che non ci sono gradi di libert\`a dinamici associati alla traccia del campo tensoriale. Con una opportuna scelta di costanti di integrazione possiamo dunque sempre ottenere una soluzione dell'equazione d'onda caratterizzata dalla condizione 
\beq
h\equiv h_\nu\,^\nu=0.
\label{a12}
\eeq

Ma se $h=0$, l'Eq. (\ref{a11}) implica $\pa^1  h_{\mu1}=0$, ovvero $h_{\mu1}=$ costante (la soluzione infatti deve dipendere da $x-ct$, per cui se $\pa^1  h_{\mu1}=0$ allora anche $\pa^0  h_{\mu 0}=0$). Dunque, come nel caso precedente, $h_{\mu 1}$ non descrive gradi di libert\`a che si propagano, ma solo contributi non dinamici che possono essere annullati,
\beq
h_{\mu 1}=0,
\label{a13}
\eeq
scegliendo opportune condizioni al contorno.

Combinando le condizioni (\ref{a10}), (\ref{a12}), (\ref{a13}) troviamo dunque che, nel sistema di coordinate che abbiamo introdotto, il campo tensoriale che descrive l'onda gravitazionale ha solo due componenti indipendenti, $h_{22}=-h_{33}$ e $h_{23}=h_{32}$, ed \`e diverso da zero solo nel piano ortogonale alla propagazione dell'onda (nel nostro caso il piano $\{x_2, x_3\}$, ossia  $\{y, z\}$). Questa conveniente scelta di coordinate viene chiamata {\em TT gauge}, ed \`e un caso particolare del {\em gauge} armonico (\ref{a7}) espresso separatamente dalle due condizioni
\beq
\pa^\nu h_{\mu\nu}=0, ~~~~~~~~~~~~~
h=0,
\label{a14}
\eeq
che impongono, rispettivamente, ``trasversalit\`a'' e ``traccia nulla''. 

In questo {\em gauge} \`e ormai consueto indicare le due componenti indipendenti dell'onda col simbolo  $h_+$ per la parte diagonale (nel nostro caso $h_+=h_{22}=-h_{33}$), e   $h_\times$ per la parte trasversa  (nel nostro caso $h_\times=h_{23}=h_{32}$). La soluzione generale dell'Eq. (\ref{a8}) pu\`o essere dunque espressa come combinazione lineare di $h_+$ e $h_\times$ introducendo due tensori di polarizzazione, $\ep^{(1)}_{\mu\nu}$ e  $\ep^{(2)}_{\mu\nu}$, tali che 
\beq
h_{\mu\nu}= h_+\ep^{(1)}_{\mu\nu}  + 
 h_\times\ep^{(2)}_{\mu\nu} .
 \label{a15}
 \eeq
Questi due tensori sono costanti, a traccia nulla, e hanno componenti non nulle solo nel piano ortogonale alla direzione di propagazione. Nel nostro caso (propagazione lungo $x$) abbiamo, in particolare,
\beq
\ep^{(1)}_{\mu\nu}= 
\begin{pmatrix}
0&0&0&0\cr
0 &0  &0 &0\cr
0 &0 &1  &0 \cr
0  &0 &0  &-1 \cr
\end{pmatrix}
, ~
\ep^{(2)}_{\mu\nu}= 
\begin{pmatrix}
0&0&0&0\cr
0 &0  &0 &0\cr
0 &0 &0  &1 \cr
0  &0 &1  &0  \cr
\end{pmatrix}.
\label{a16}
\eeq
Questi due tensori soddisfano la relazione di ``ortonormalit\`a'' 
$\ep^{(i)}_{\mu\nu}\ep^{(j)\mu\nu} = 2 \da^{ij}$, con $i,j=1,2$, e quindi descrivono due stati di polarizzazione linearmente indipendenti.

Come nel caso delle onde e.m., anche in questo caso possiamo definire gli stati di polarizzazione circolare mediante opportune combinazioni  (con coefficienti complessi) degli stati di polarizzazione lineari. I corrispondenti tensori di polarizzazione circolare, per le onde gravitazionali, sono  definiti come
\beq
\ep^{(\pm)}_{\mu\nu}={1\over 2} \left[ \ep^{(1)}_{\mu\nu} \pm i \ep^{(2)}_{\mu\nu} \right],
\label{a17}
\eeq
e sono anch'essi ortonormali come quelli lineari. Le loro propriet\`a di trasformazione rispetto alle rotazioni permettono di determinare {\em l'elicit\`a} $h$ tipica dell'onda, ossia la proiezione  del momento angolare intrinseco 
$\vec{s}$ lungo la direzione di propagazione, individuata dalla  quantit\`a di moto $\vec p$:  in unit\`a di Planck, $h=\vec{s} \cdot \vec p/|{\vec p}|$.

Ricordiamo a questo proposito che, data un'onda $\psi$ polarizzata circolarmente che si propaga in direzione $\wh n$, si dice che l'onda ha elicit\`a $h$ se, sotto una rotazione di un angolo $\theta$ attorno alla sua direzione di moto, l'onda si trasforma come segue:
\beq
\psi \ra \psi'=  \psi\, e^{ih \theta} .
\label{a18}
\eeq
Prendiamo dunque un'onda gravitazionale che si propaga lungo l'asse $x$, caratterizzata dai tensori di polarizzazione circolare (\ref{a17}), e consideriamo una rotazione di un angolo $\theta$ nel piano $\{y, z\}$, 
descritta dalla matrice
\beq
R_\mu\\^\a= 
\begin{pmatrix}
1&0&0&0\cr
0 &1  &0 &0\cr
0 &0 &\cos \theta  &\sin \theta \cr
0  &0 &- \sin\theta  &\cos \theta \cr
\end{pmatrix} . 
\label{a19}
\eeq
Utilizzando la rappresentazione esplicita dei tensori di polarizzazione si trova facilmente che
\beq
\ep^{\pr (\pm)}_{\mu\nu} = 
R_\mu\\^\a R_\nu\\^\b \ep^{(\pm)}_{\a\b}
=\ep^{(\pm)}_{\mu\nu} \,e^{\pm 2 i \theta} .
\label{a20}
\eeq
In confronto con l'Eq. (\ref{a18}) ci dice immediatamente che l'onda gravitazionale ha elicit\`a $h= \pm 2$.

\section{Appendice 2. L'energia irraggiata nell'approssimazione quadrupolare} 

\`E ben noto che la radiazione e.m. emessa da un sistema di cariche non relativistiche \`e controllata, in prima approssimazione, dalla derivata temporale seconda del momento di dipolo elettrico, $\ddot{\vec d}$, dove $\vec d= \sum_i q_i  {\vec x}_i$ (abbiamo supposto, per semplicit\`a, che il sistema sia composto da cariche $q_i$ puntiformi). La potenza irraggiata $dE/dt$, in questa approssimazione, \`e data da
\beq
{dE\over dt}= {2\over 3 c^3} \left|  \ddot{\vec d} \right|^2.
\lab{b1}
\eeq

Nel caso della gravit\`a non ci pu\`o essere radiazione di questo tipo: il corrispondente dipolo gravitazionale deve avere derivata seconda nulla per la conservazione della quantit\`a di moto totale (si veda l'esempio presentato nel seguito di questa Appendice,  e in particolare le equazioni (\ref{b4}), (\ref{b5})).
Per avere un flusso di radiazione diverso da zero, a distanze arbitrariamente grandi dalle masse che fanno da sorgenti, dobbiamo spingerci oltre l'approssimazione dipolare, e calcolare le fluttuazioni della geometria (ossia, il tensore $h_{\mu\nu}$ definito nell'App. 1) prodotte dal cosiddetto {\em momento di quadrupolo} delle sorgenti, rappresentato dal tensore $Q_{\mu\nu}$.

Se lavoriamo nel {\em gauge TT} (definito nell'Appendice 1, si veda l'Eq. (\ref{a14})) 
 \`e sufficiente limitarci alle componenti spaziali di questo tensore, che sono definite da
\beq
Q_{ij}= \int d^3 x \, \r (x,t) \left(3 x_i x_j- |\vec x|^2 \da_{ij}\right),
\label{b2}
\eeq
dove $\r$ \`e la densit\`a di massa del sistema. L'energia della radiazione gravitazionale emessa per unit\`a di tempo, in questo caso, \`e controllata dalla derivata temporale terza del momento di quadrupolo, $ \stackrel{\dots}{Q}_{ij}$, ed \`e data da
\beq
{dE\over dt}=  - {G\over 45  c^5}  \stackrel{\dots}{Q}_{ij}\stackrel{\dots}{Q}^{ij} ,
\label{b3}
\eeq
dove $G$ \`e la costante di Newton (si veda ad esempio \cite{4}, Cap. 9.2.3).

Per illustrare questo risultato, e capirne meglio alcune sue implicazioni, 
\`e conveniente concentrarsi su un semplice ma realistico esempio. Consideriamo un sistema binario composto da due masse puntiformi $m_1$ e $m_2$, con coordinate $\vec  x_1$ e $\vec x_2$, che orbitano sotto l'azione della reciproca attrazione gravitazionale muovendosi con velocit\`a non relativistiche. Il loro centro di massa \`e localizzato del punto di coordinate 
\beq
\vec X_{CM}= {m_1 \vec  x_1+ m_2 \vec  x_2 \over m_1+m_2},
\lab{b4}
\eeq
e si muove di moto rettilineo e uniforme, $\ddot{\vec X}_{CM}=0$. Il corrispondente dipolo gravitazionale del sistema,
\beq
\vec d_g= m_1 \vec  x_1+ m_2 \vec  x_2 = (m_1+m_2)\vec X_{CM},
\lab{b5}
\eeq
soddisfa dunque la condizione  $\ddot{\vec d}_g=0$, 
per cui non viene emessa radiazione gravitazionale di tipo dipolare (come anticipato all'inizio di questa Appendice). 

Per calcolare il momento di quadrupolo \`e conveniente rappresentare la dinamica del nostro sistema come il moto di un singolo corpo puntiforme di massa ridotta $M$,
\beq
M= {m_1m_2\over m_1+m_2},
\label{b6}
\eeq
che orbita attorno alla traiettoria del centro di massa percorrendo un'orbita chiusa, che supponiamo localizzata nel piano $\{x_1, x_2\}$. Supponiamo anche, per semplicit\`a, che tale traiettoria sia una circonferenza di raggio $a=$ costante, percorsa con frequenza orbitale $\om$ , e quindi descritta dalle equazioni parametriche
\beq
x_1= a \cos \om t, ~~~~~~~~ x_2= a \sin \om t, ~~~~~~~~ x_3=0.
\label{b7}
\eeq

Poich\'e il corpo \`e puntiforme, la sua densit\`a di massa si esprime con la delta di Dirac, 
\beq
\r= M \da (x_1- a \cos \om t) \da (x_2 - a \sin \om t) \da (x_3),
\label{b8}
\eeq
e la definizione (\ref{b2}) fornisce immediatamente le seguenti componenti del tensore di quadrupolo:
\bea
&&
Q_{11}= M a^2 \left( 3 \cos^2 \om t -1\right), ~~~~ Q_{33}= -M a^2, 
\nonumber \\ &&
Q_{22}= M a^2 \left( 3 \sin^2 \om t -1\right),
\nonumber \\ &&
Q_{12}= Q_{21}=3 M a^2 \cos \om t \sin \om t.
\label{b9}
\eea
A questo punto diventa un semplice esercizio calcolarne la derivata temporale terza e inserirla nella definizione (\ref{b3}) per ottenere la potenza totale emessa. 

Questo calcolo esplicito porta a un risultato fortemente oscillante nel tempo, con un'ampiezza modulata dalla frequenza orbitale $\om$. Se ci interessa studiare la radiazione emessa dal sistema binario su tempi molto pi\`u lunghi del suo periodo, $T=2\pi/\om$, ci conviene calcolare la potenza mediata su un periodo,  $\left\langle{dE/ dt}\right\rangle$, 
ottenuta integrando sul tempo da $0$ a $T$, ossia ponendo
\bea
&&
\left\langle{dE\over dt}\right\rangle=  - {G\over 45  c^5}\l  \stackrel{\dots}{Q}_{ij}\stackrel{\dots}{Q}^{ij} \re 
\nonumber \\ &&
\equiv   - {G\over 45  c^5}{1\over T} \int_0^T dt  \stackrel{\dots}{Q}_{ij}\stackrel{\dots}{Q}^{ij}. 
\label{b10}
\eea
Una semplice integrazione fornisce allora il risultato
\beq
\left\l {dE\over dt}\right \re =
-{32 G\over 5 c^5} \,M^2 a^4 \om^6 ,
\label{b11}
\eeq
gi\`a anticipato qualitativamente nella Sezione 1.

Il fattore costante $G/c^5 $ che controlla l'intensit\`a della radiazione gravitazionale di quadrupolo (e che vale, in unit\`a c.g.s., $G/c^5 \sim 10^{-60}$) ci fa capire l'estrema debolezza della radiazione, a meno che non si tratti di sorgenti di grandi dimensioni (di tipo astrofisico). Ad esempio, se prendiamo una massa pari a quella solare, $M \sim 10^{33}$ g, un raggio orbitale pari a dieci raggi solari, $a \sim 10^{11}$ cm, e un periodo orbitale di qualche ora, $\om \sim 10^{-4}$ Hz, l'Eq. (\ref{b11}) fornisce una potenza media di circa $10^{20}$ Watt.

Il flusso di energia ricevuto sulla Terra, per\`o, si riduce in modo inversamente proporzionale al quadrato della distanza. Supponiamo, ad esempio, che la sorgente si trovi all'interno della nostra galassia, a una distanza tipica dell'ordine di $R \sim 10^{20}$ cm. Il flusso di energia ricevuto sulla Terra (usando per 
$\left\langle{dE/ dt}\right\rangle$ 
il valore numerico precedente) diventa, in questo caso, 
\bea
\nonumber
\Phi &=& 
{1\over 4 \pi R^2} \left| \left \l {dE \over dt}\right\re\right| 
\simeq10^{-14} \frac{\textrm{erg} } { \textrm{cm}^2 \textrm{sec} }
\\
& \simeq &10^{-21} \textrm{ Watt} \over \textrm{cm}^2.
\label{b12}
\eea

\WideFigure{f8}
{\label{fig8}Risposta al modo di polarizzazione $h_+$ per masse di prova disposte in cerchio nel piano ortogonale al moto dell'onda.}

\section{Appendice 3. La risposta delle masse al passaggio di un'onda} 

Sappiamo che il passaggio di un'onda e.m. fa vibrare le cariche elettriche che incontra al suo passaggio. Allo stesso modo, ci aspettiamo che un'onda gravitazionale faccia vibrare le masse. Ma in che modo?

La risposta delle masse \`e descritta dalla cosiddetta equazione di ``deviazione geodetica'', che si scrive
\beq
A^\mu \equiv{D^2 \eta^\mu \over d\tau^2}=- \eta^\nu R_{\nu\a\b}\\^\mu u^\a u^\b,
\label{c1}
\eeq
e che fornisce la relativa accelerazione covariante $A^\mu$ tra due masse di prova che si muovono con 4-velocit\`a $u^\mu$ e hanno separazione spaziale descritta dal 4-vettore $\eta^\mu$. La variabile $\tau$ \`e il tempo proprio, e $R$ indica il tensore di Riemann.

Consideriamo due masse inizialmente a riposo, separate da una distanza spaziale $\eta^\mu= (0, L^i)$, dove $L^i=$ costante. Sotto l'azione dell'onda si mettono in moto, e la loro distanza relativa cambia in funzione del tempo. Se il campo gravitazionale dell'onda \`e sufficientemente debole possiamo supporre che il moto delle masse sia non-relativistico, e che i loro spostamenti siano piccoli. In questa approssimazione poniamo
\beq
\eta^\mu= L^\mu + \xi^\mu, ~~~~~~~~~ |\xi| \ll |L|, 
\label{c2}
\eeq
approssimiamo la velocit\`a con $u^\mu= (c, \vec 0)$, e ci limitiamo a termini del primo ordine in $\xi$ e nelle componenti del tensore $h_{\mu\nu}$ che descrive l'onda. L'Eq. (\ref{c1}) si riduce a 
\beq
\ddot \xi^i= -L^j R_{j00}\\^i c^2,
\label{c3}
\eeq
dove il punto indica la derivata rispetto al tempo, e il tensore di Riemann \`e calcolato al primo ordine in $h$ utilizzando la connessione (\ref{a4}). Lavorando nel {\em gauge TT} dove $h_{\mu0}=0$ (si veda App.1) si ottiene
 \beq
 R_{j00}\\^i={1\over 2c^2} \ddot h_j\\^i,
\label{c4}
\eeq
e quindi
\beq
\ddot \xi^i= -{1\over 2}   \ddot h_j\\^i L^j.
\label{c5}
\eeq

Prendiamo per semplicit\`a un'onda piana monocromatica, che si propaga lungo l'asse $x_1$ con frequenza $\om=ck$. Le sue componenti non nulle sono confinate nel piano $\{x_2, x_3\}$, e sono date da:
\bea
&&
h_{22}=-h_{33}=h_+ \cos \left(kx-\om t \right),
\nonumber \\ &&
h_{23}=h_{32}=h_\times \cos \left(kx-\om t \right),
\lab{c6}
\eea
dove $h_+$ e $h_\times$ sono costanti. 
Abbiamo quindi $\ddot h_{ij}= -\om^2 h_{ij}$, e lo spostamento delle masse nel piano perpendicolare al moto dell'onda \`e descritto dalle equazioni
\bea
&&
\!\!\!\!\!\!\!\!\!\!\!\!
\ddot \xi^2=-{\om^2\over 2} \left( h_+ L_2+ h_\times L_3\right) \cos \left(kx-\om t\right),
\nonumber \\ &&
\!\!\!\!\!\!\!\!\!\!\!\!
\ddot \xi^3=-{\om^2\over 2} \left( h_\times L_2-  h_+L_3 \right) \cos \left(kx-\om t\right).
\label{c7}
\eea

Per illustrare meglio gli effetti dell'onda supponiamo ora di avere un insieme di masse di prova che a riposo sono disposte in modo da formare un cerchio di raggio $L/2$ nel piano $\{x_2, x_3\}$: possiamo dunque porre, nell'equazione precedente, $L_2=L_3=L$. Supponiamo anche che l'onda incidente sia polarizzata linearmente,  con ampiezza $h_+\not=0$ e  $h_\times=0$. 

In questo caso, la forza (\ref{c7}) che agisce sulle masse di prova varia periodicamente, passando da una fase con $\cos \left(kx-\om t\right)=1$, caratterizzata da
\beq
\ddot \xi^2=-{\om^2\over 2}  h_+ L, ~~~~~~~~~~~
\ddot \xi^3={\om^2\over 2} h_+ L,
\label{c8}
\eeq
(ovvero attrazione massima lungo $x_2$ e repulsione massima lungo $x_3$), ad una fase con $\cos \left(kx-\om t\right)=-1$, caratterizzata da
\beq
\ddot \xi^2={\om^2\over 2}  h_+ L, ~~~~~~~~~~~
\ddot \xi^3=-{\om^2\over 2} h_+ L,
\label{c8}
\eeq
(ovvero repulsione massima lungo $x_2$ e attrazione massima lungo $x_3$). Seguendo la variazione periodica dell'ampiezza di $h_+$ il cerchio di masse subisce quindi una serie di successive e alternate compressioni/espansioni lungo gli assi ortogonali $x_2$ e $x_3$, come illustrato in Fig. \ref{fig8}.

\WideFigure{f9}
{\label{fig9}Risposta al modo di polarizzazione $h_\times$ per masse di prova disposte in cerchio nel piano ortogonale al moto dell'onda.}

Se l'onda \`e polarizzata in modo opposto ($h_+=0$, $h_\times \not= 0$), la situazione \`e simile. In questo secondo caso, infatti, le equazioni del moto per le masse di prova che otteniamo dalla (\ref{c7}) differiscono dalle precedenti solo per una rotazione di un angolo $\pi/4$ nel piano $\{x_2, x_3\}$. L'effetto del modo $h_\times$ sul cerchio di masse \`e dunque lo stesso del modo $h_+$, ma \`e riferito a un sistema di assi ruotato di 45 gradi rispetto alla precedente configurazione, come illustrato nella Fig. \ref{fig9}.

I due tipi di distorsione (o ``{\em stress}'') indotti sulle masse di prova e illustrati in figura sono tipici degli stati di polarizzazione di un'onda di tipo tensoriale. L'obiettivo dei rivelatori \`e concettualmente semplice (ma tecnicamente molto complicato): amplificare al massimo tali distorsioni, riducendo contemporaneamente al minimo tutti i possibili ``rumori'' estranei, ossia tutti gli effetti (termici, sismici, etc) che possono produrre ulteriori vibrazioni delle masse di prova, e che non sono dovuti direttamente al passaggio dell'onda gravitazionale.

\section{Ringraziamenti}
Alcune parti di questo articolo sono basate su lavori effettuati col parziale supporto del MIUR (progetto PRIN) e dell'INFN (programma TAsP).


\AuthorsBio{Maurizio Gasperini}{\`e attualmente professore ordinario di Fisica Teorica all'Universit\`a di Bari, dove svolge attivit\`a scientifica e didattica nel campo della relativit\`a, della cosmologia, della gravitazione e della teoria delle interazioni fondamentali. \`E stato in precedenza ricercatore all'Universit\`a di Torino, {\em Academic Staff Member} alla Universit\`a  della California (Santa Barbara) e {\em Scientific Associate} al CERN dove, in collaborazione con Gabriele Veneziano, ha formulato e sviluppato un modello di universo primordiale basato sulla teoria delle stringhe. Per ulteriori dettagli si veda 
\url{http://www.ba.infn.it/~gasperin/academic.html}}

\end{document}